# Triplet-Pair Spin Signatures from Macroscopically Aligned Heteroacenes in an Oriented Single Crystal

Brandon K. Rugg,[1] Kori E. Smyser,[2,3] Brian Fluegel,[1] Christopher H. Chang,[1,2] Karl J. Thorley,[4] Sean Parkin,[5] John E. Anthony,[4,5] Joel D. Eaves,[2,3] Justin C. Johnson[1,2*]

[1]Chemistry and Nanosciences Center, National Renewable Energy Laboratory, 15013 Denver West Parkway, Golden, Colorado 80401, USA

[2]Renewable and Sustainable Energy Institute, University of Colorado Boulder, Boulder, Colorado 80309, USA

[3]Department of Chemistry, University of Colorado, Boulder, Colorado 80309, USA

[4]University of Kentucky Center for Applied Energy Research, Lexington, Kentucky 40511, USA

[5]Department of Chemistry, University of Kentucky, Lexington, Kentucky 40506, USA

*Corresponding author: Justin C. Johnson

**Email:**  justin.johnson@nrel.gov

**Author Contributions:** B.K.R prepared samples and performed TR-EPR measurements and analysis, under advisement of J.C.J.  C.H.C. performed density functional theory calculations. B.F. performed magnetoluminescence experiments.  K.J.T. synthesized TES TIPS-TT, under advisement of J.E.A.  S.P. characterized and indexed single crystals.  K.E.S. and J.D.E developed theoretical methods and performed TR-EPR simulations.  B.K.R., J.C.J., K.E.S, and J.D.E. wrote the manuscript with input from all authors.

**Abstract**

The photo-driven process of singlet fission generates coupled triplet pairs (TT) with fundamentally intriguing and potentially useful properties. The quintet $^5TT_0$ sublevel is particularly interesting for quantum information because it is highly entangled, addressable with microwave pulses, and could be detected using optical techniques. Previous theoretical work on a model Hamiltonian and nonadiabatic transition theory, called the *JDE* model, has determined that this sublevel can be selectively populated if certain conditions are met. Among the most challenging, the molecules within the dimer undergoing singlet fission must have their principal magnetic axes parallel to one another and to an applied Zeeman field. Here, we present time-resolved paramagnetic resonance (TR-EPR) spectroscopy of a single crystal sample of a novel tetracenethiophene compound featuring arrays of dimers aligned in this manner, mounted so that the orientation of the field relative to the molecular axes could be controlled. The observed spin sublevel populations in the paired TT and unpaired (T+T) triplets are consistent with predictions from the *JDE* model, including preferential $^5TT_0$ formation at $z \parallel B_0$, with one caveat—two $^5TT$ spin sublevels have little to no population. This may be due to crossings between the $^5TT$ and $^3TT$ manifolds in the field range investigated by TR-EPR, consistent with the inter-triplet exchange energy determined by monitoring photoluminescence at varying magnetic fields.

**Significance Statement**

Producing ordered elements of quantum information, such as the oriented spin of an electron, is challenging at large scales and reasonable temperatures. Molecules crystallized into arrays that possess oriented spins upon photoexcitation can surmount such challenges. We demonstrate the design and characterization of a tailored molecule that preferentially aligns with macroscopic order and possesses photoexcited species with characteristic spin signatures that depend on the orientation of the crystal with respect to a magnetic field. We rationalize the behavior with a rigorous model that explains the primary electron paramagnetic resonance (EPR) features and leads toward crucial design principles for further achieving the desired outcomes.

**Main Text**

**Introduction**

Technologies that utilize quantum information (QI) have the potential to transform the fields of computation, sensing, and communications. Still, such applications are currently beyond reach, in large part because of difficulties that prevent sufficient scaling of conventional materials architectures currently used to store registers of qubits, or units of QI.(1, 2) However, the "bottom-up" approach of designing, synthesizing, and optimizing molecules to serve as spin qubit candidates has emerged as a promising alternative, as molecular qubits can achieve long coherence times, are amenable to organization in extended qubit structures in a controlled fashion, and enable new varieties of photophysical control.(3, 4)

Among molecular candidates, singlet fission (SF) materials are worthy of attention as they can form pure, entangled quantum states involving two triplet excitons (T) upon photoexcitation, even at room temperature.(5-8) In addition to the possibility of selective state population, the two-exciton states resulting from SF have already shown near-microsecond coherence times well above milliKelvin temperatures required for other systems, enabling spin manipulation and characterization via pulsed electron paramagnetic resonance (EPR) experiments.(9, 10) As an example, it has been demonstrated that the polarization of the exciton pair TT can be transferred to specific nuclei on the same molecule using controlled microwave pulses,(9) which may allow for a hybrid qubit that takes advantage of the higher polarization and faster manipulability of electron spins and the much longer coherence times of nuclear spins.(11) Further, optical readout of the spin state is inherent to many molecular qubits.(12-14)

The spin-conserving evolution of an excited singlet exciton ($S_1 + S_0$) on a chromophore pair to the overall singlet $^1TT$, delocalized over both chromophores, is the first step of SF.(15) In addition to the $^1TT$ state, there are three triplet $^3TT$ and five quintet $^5TT$ sublevels, split from $^1TT$ by $J$ and $3J$, respectively, where $J$ is the intertriplet exchange interaction. The relative ordering of the spin state manifolds is dictated by the sign of $J$, with a negative, or ferromagnetic $J$ indicating that the $^5TT$ state is lowest in energy.

Smyser and Eaves previously developed a model, based on nonadiabatic transition theory (NTT), for dimers whose molecules share principal axis directions.(6) In this "*JDE* model," the effective *J* is large enough to separate the $^{2S+1}TT$ states, but, immediately following singlet fission, large fluctuations in *J* induce crossings between the various $^{2S+1}TT_M$ sublevels to facilitate relaxation events.(16) The subsequent sublevel population is dictated by the orientation of the molecular z-axis relative to an applied magnetic field ($B_0$). Importantly, for quantum information applications, the $^5TT_0$ sublevel is dominant for $z \parallel B_0$, and there is evidence that it is addressable with microwave pulses and has the potential for optical readout.(14, 17, 18) In this paper, we expand the model to include exciton unbinding dynamics that can occur in crystals with mobile excitons and find that the separated triplets (T+T) maintain the spin polarization of the initially formed sublevels.

Motivated by the predictions of the *JDE* model, we have conducted a time-resolved EPR (TR-EPR) study of a single crystal of 2-triethylsilyl-5,11-bis(triisopropylsilyl ethynyl)

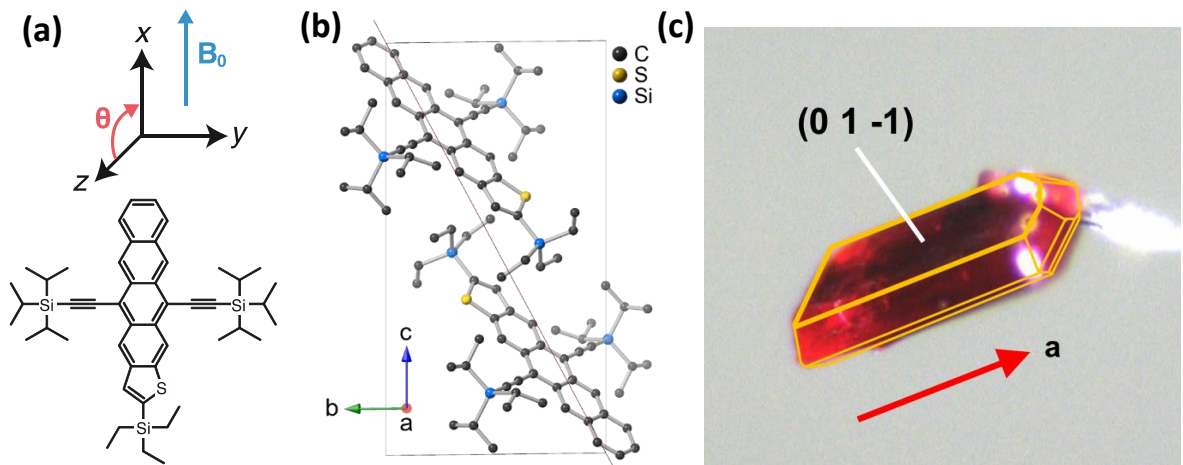

*Figure 1.* TES TIPS-TT molecular and crystal structure. (a) Molecular coordinate system. The primary axis z is perpendicular to the molecular π-system and its orientation relative to an applied magnetic field ($B_0$) is defined by $\theta$. The illustration shows $\theta = 90°$, or $x \parallel B_0$. (b) Crystal structure of TES TIPSTT from the view along the a-axis, with the (0 1 -1) face indicated by a red line. The S atoms in both molecules appear on either side of the thiophene ring with equal likelihood. H atoms are hidden. (c) Photograph of a representative TES TIPSTT crystal, which typically forms a tablet-shaped monolith with its major face aligned with the (0 1 -1) plane and its long axis equivalent to the a-axis of the crystal structure.

tetraceno[2,3-b]thiophene (TES TIPS-TT), a novel heteroacene with a crystal structure in which all molecules share a common *z*-axis (Figure 1). The macroscopic properties of the TES TIPS-TT crystal also permit samples to be prepared so that the angle (θ) of the molecular *z*-axis relative to $B_0$ can be systematically controlled. This approach allows for an unprecedented look into the orientation dependence of spin evolution relative to previous studies of $^5$TT, which have employed disordered samples or have at best achieved partial ordering.(10, 19-21) Other single crystal studies of SF materials, such as one performed on tetracene, did not directly observe TR-EPR signal from $^5$TT.(22) Sublevel populations of photoexcited TES TIPS-TT triplet pairs are well-described by the *JDE* model. However, primarily absorptive or anomalously broadened TR-EPR features are also observed that require a more detailed understanding of dynamics in this crystalline material. Nonetheless, routes toward selective population of $^5$TT$_0$ are outlined, laying the groundwork for further studies of the quantum properties of this state.

**Results**

**Structural characterization.** Knowledge of the microscopic and macroscopic crystal properties of TES TIPS-TT 

described by the *JDE* model. However, primarily absorptive or anomalously broadened TR-EPR features are also observed that require a more detailed understanding of dynamics in this crystalline material. Nonetheless, routes toward selective population of $^5TT_0$ are outlined, laying the groundwork for further studies of the quantum properties of this state.b-c) enables a high level of control over molecular orientation relative to $B_0$ in TR-EPR experiments. The unit cell features two unique inversion-related molecules of TES TIPS-TT that are defined by a common molecular *z*-axis, which simplifies both the spin dynamics of the SF exciton pair and interpretation of the associated TR-EPR spectra relative to an unaligned chromophore pair.(6) We note that prior work on a series of related tetraceno-bithiophene derivatives discovered high mobility in thin-film transistors and related this behavior to the slip-stacked packing found in single crystals, suggesting strong and extended π-π interactions.(23) Here, the substitution of an alkylsilyl group on the thiophene ring enforces asymmetry that renders primarily distinct dimer types into the structure. In addition to the dimer observable in the unit cell that has the thiophene rings partially eclipsed (Figure 1b), each chromophore is also similarly coupled to another neighboring molecule, but with the thiophenes on opposite sides and the terminal phenyls partially eclipsed (Figure 2, dimer II). This series of stacked dimers continues in a staircase-like fashion. Perpendicular to this direction, there are two varieties of "side-by-side" dimers, which appear to be much more weakly coupled (Figure 2, dimer III). While not likely to be the dominant sites for singlet fission, due to the lack of strong orbital overlap, these nearly co-planar dimers may play a role in supporting weakly coupled triplet pairs upon diffusion and nongeminate encounters.

TES TIPS-TT forms tablet-shaped crystals with a clearly identifiable long axis and a parallel set of two large faces (Figure 1c). The results obtained from indexing several crystals indicate that the long axis corresponds to the crystallographic *a*-axis, and the large faces correspond to the (0 1 -1) and (0 -1 1) planes of the unit cell. Combining this information with the orientation of the molecules within the unit cell, it is determined that the molecular *z*-axes of all chromophores within a single-crystal sample could be aligned with $B_0$ by mounting the crystal to quartz rods cut at 38°, as shown in Figure S6. Rotation of the EPR sample rod permits careful control of the orientation of both the *x*- and *z*-axes relative to $B_0$, while the *y*-axis remains perpendicular.

**Calculations.** Table 1 shows calculated exchange couplings for four dimer models (Figure 2). Dimers Ia and Ib are nearly identical and only differ in whether sulfur positions were modeled on the same or opposite sides, respectively, of a plane encompassing the long molecular axes of the dimer pair.

**Table 1.** SCAN/def2-TZVPP calculated exchange couplings for TES TIPS-TT dimers.

| Dimer model | Calculated exchange coupling (GHz) |
|---|---|
| Ia | -15 |
| Ib | 35 |
| II | -315 |
| III | < 1 |

Dimers Ia and Ib have similar exchange coupling magnitudes, but of opposite sign. When the sulfur atoms are on opposite sides of the aromatic core (Ib) the quintet state is higher in energy than the singlet state (i.e., antiferromagnetic coupling). Dimer model Ia shows a ferromagnetic coupling of comparable magnitude to that observed experimentally (*vide infra*). The primary interaction for these dimers is through the thiophene rings. In contrast, dimer II involves the interaction between the distal tetracene ends. Both enhanced spatial overlap and intermolecular proximity may explain the larger calculated exchange coupling for dimer II, as the molecular planes are separated by 3.3 Å, compared to 5.1 Å for Ia/b. Dimer III involves side-by-side molecules and possesses minimal exchange splitting between the broken-symmetry singlet and quintet states (the raw energy difference was calculated to be -0.018 GHz).

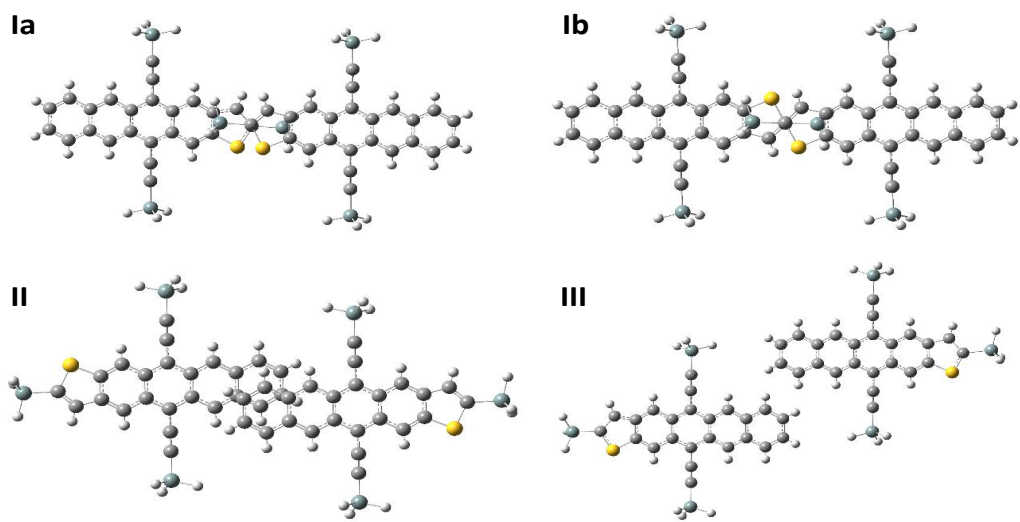

*Figure 2. Dimer pairs for which exchange couplings were calculated. Dimers Ia and Ib differ in the modeled disorder of sulfur atoms in the crystal structure. Images have phenyl rings in the plane of the page.*

**Magnetophotoluminescence.** Single crystals of TES TIPS-TT exhibit several bands of steady-state fluorescence in the range of 600-800 nm when excited at 519 nm (Figure S7). The yield of the fluorescence within the range of 700-775 nm shows a clear dependence on the strength of an applied magnetic field at low temperatures (Figure 3a). Dips in the fluorescence intensity are observed where the non-magnetic $^1$TT state crosses with the dark $^5$TT$_M$ sublevels that tune through the magnetic Zeeman interaction.(17) The magnitude (though not sign) of $J$ can be determined from the distribution of the dips in the spectrum. The experimental field range (0 - 14 T) allows for detection of $J$ between roughly 5 and 131 GHz. As it appears to be most consistent with the DFT calculations and TR-EPR spectra (*vide infra*), we assume ferromagnetic coupling, as shown in the Figure 3b.

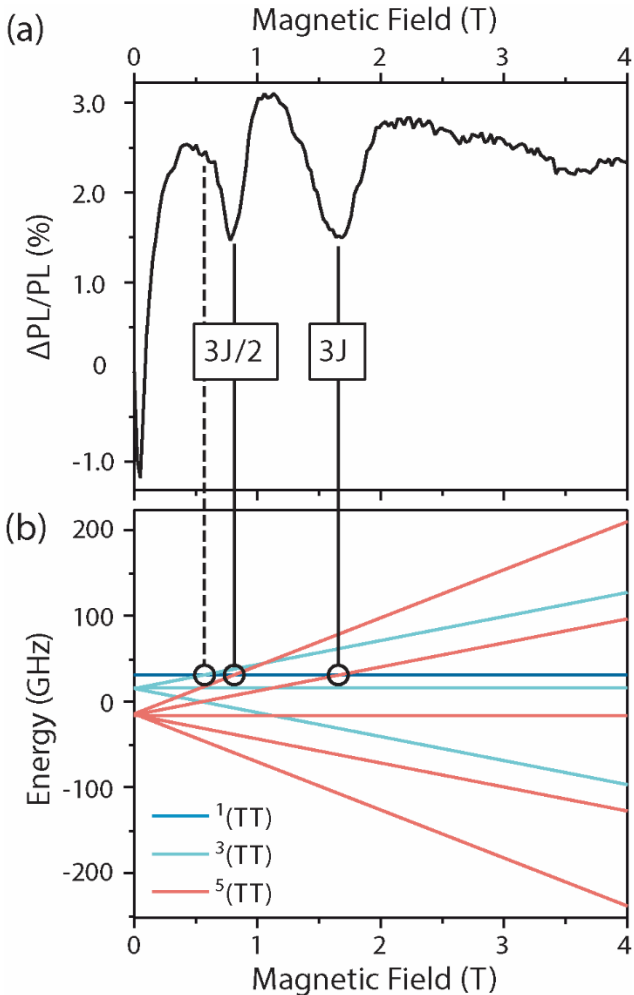

**Figure 3.** Magnetophotoluminescence of TES TIPS-TT. (a) Changes in relative PL vs. applied magnetic field at 2 K. (b) Energy level diagram of an exciton pair with $J$ = -15.4 GHz in a magnetic field. Black circles highlight level crossings between $^1$TT and $^5$TT$_{+1,+2}$ (observed) and with $^3$TT$_{+1}$ (not observed).

Two prominent dips in the field sweep appear at 0.83 and 1.65 T. Based on the 1:2 ratio of these values, and assuming $J > 0$, the first and second peaks can be assigned to $^1$TT mixing with $^5$TT$_{+2}$ and $^5$TT$_{+1}$ (Figure 3b). This occurs at field strengths of about 1.5|$J$| and 3|$J$|, respectively, so that |$J$| = 15.4 ± 0.3 GHz for at least one dimer within the TES TIPS-TT crystal

structure. No dip is detected at the $^1TT - {}^3TT_{+1}$ curve crossing field position, (at $|J|$, dashed line in Figure 3). Its absence here is not related to excessive broadening, as a peak at $J$ should be narrower than higher-field features.(18)  A dip and subsequent rise is observed at low field, but no other features are found up to 14 T (Figure S8).

These results are similar to those previously obtained from TIPS tetracene,(17) which we have reproduced using a single crystal sample to verify the instrument sensitivity (Figure S9). However, three peaks (rather than just two) were observed for TIPS tetracene. In addition to the peaks consistent with $^5TT_{\pm 2}$, and $^5TT_{\pm 1}$ mixing with $^1TT$, there is also a peak at $J$, suggesting mixing with $^3TT_{\pm 1}$. This yields a characteristic 1:3/2:3 splitting pattern that was observed by Bayliss *et al*. for multiple dimer sites within semi-crystalline samples of TIPS tetracene.(17)

**Electron paramagnetic resonance spectroscopy.** TR-EPR spectra of TES TIPS-TT in a 4:1 mixture of iodobutane and toluene after $\lambda_{ex}$ = 600 nm were collected at 100 K (Figure S10). The heavy atom effect from the solvent encourages intersystem crossing (ISC) in the solute, allowing the triplet state of isolated TES TIPS-TT molecules to be characterized. Fitting the extracted spectrum indicated zero-field splitting (ZFS) parameters of $D$ = 1273 MHz and $E$ = -40 MHz. These parameters represent the axial ($z$) and transversal ($x$, $y$) components of the spin dipole-dipole interaction between the two electron spins within a single triplet species (T). They also inform on the field range for the $^3TT$, $^5TT$, and T+T spectral features in EPR a from crystalline TES TIPS-TT. The two prominent peaks observed in the triplet powder spectrum (e.g. around 325 and 370 mT) are associated with the statistically favored $z \perp B_0$ orientation and are split by $|D - 3E|$.

TR-EPR spectra obtained from a crystalline powder of TES TIPS-TT at room temperature after $\lambda_{ex}$ = 610 nm are shown below the crystal spectra in Figure 4 and are expanded in Figure S10 for comparison with the solvated molecule spectra. Qualitatively, the presence of $^5TT_0$ in the

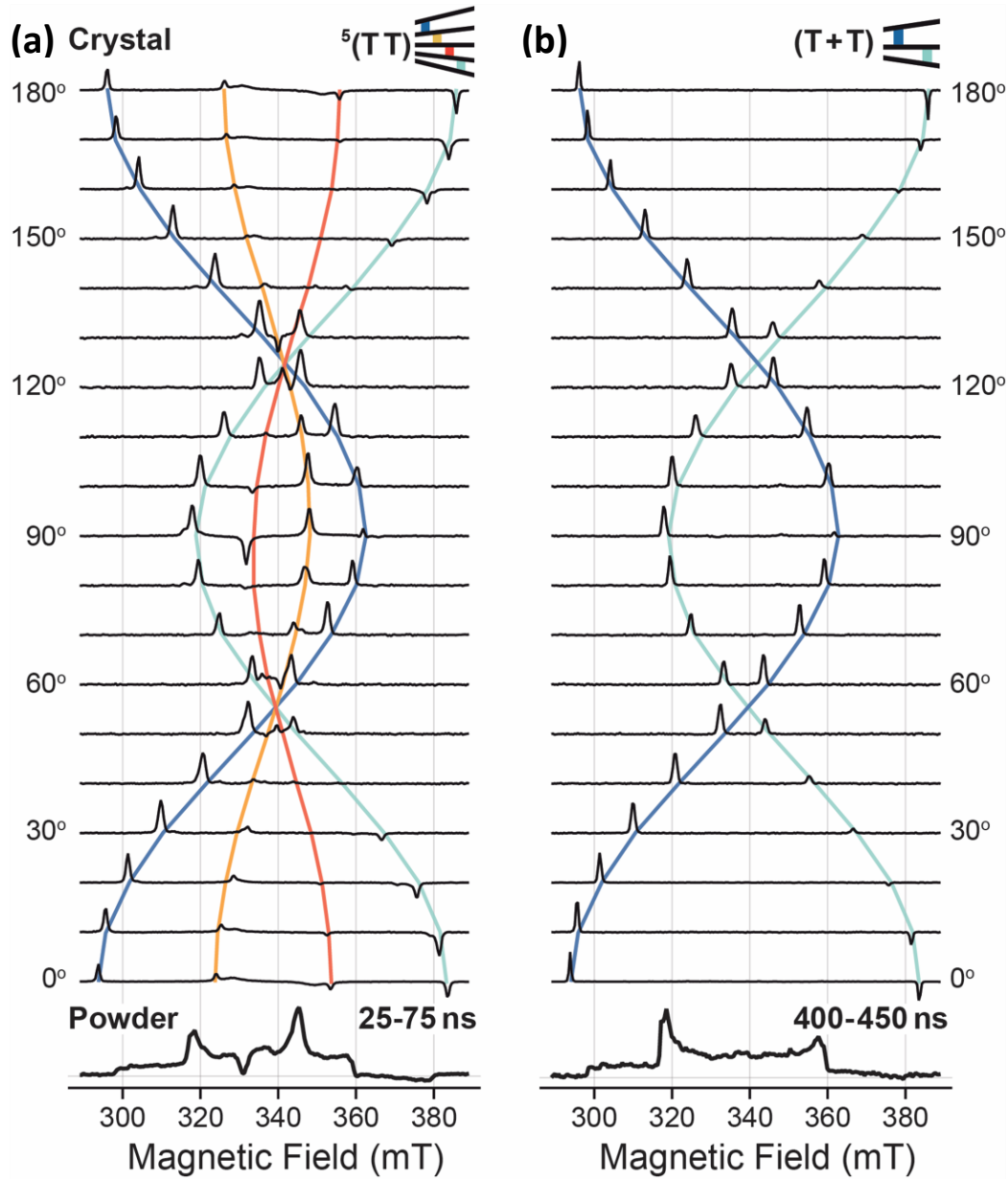

*Figure 4.* TR-EPR spectra of a single crystal of TES TIPS-TT mounted to enable rotation of the molecular x- and z-axes in the plane of $B_0$ at (a) early (25 – 75 ns) and (b) late (400 – 450 ns) times. The starting orientation of 0° represents the orientation in which z || $B_0$. Transitions within $^5TT$ and T+T are color-coded with identical colors being used for transitions of the same energy. Powder spectra obtained at the respective time ranges are shown at bottom.

early time 25 – 75 ns spectrum can be inferred from the presence of a pair of peaks, one emissive and one absorptive, split from each other about center field by approximately $|D|/3$.(10, 20) Other features in the spectrum, that are primarily absorptive, persist to later times (400 – 450 ns), after the most characteristic $^5TT_0$ signatures have disappeared, and are likely associated with T+T. State relaxation for strongly coupled and well-aligned dimers is expected to populate $^5TT_{\pm M}$ sublevels symmetrically, e.g., population of $^5TT_{+2}$ should result in equal population of $^5TT_{-2}$.(6) However, preferential populations of the lower $M$ sublevels have been observed previously.(24-26)

A single crystal of TES TIPS-TT was mounted to make the orientation $z \parallel B_0$ attainable within the EPR spectrometer. Starting with $z \parallel B_0$ (labeled 0°), the sample was rotated to collect TR-EPR spectra for different orientations about the $y$-axis in 10° increments between 0 and 180° (Figure 4 and Figures S11-S13). Figure 4a shows lines corresponding to $^5TT$ transitions for strongly coupled dimers (i.e. $|J|$ is sufficiently large so that the $^{2S+1}TT$ manifolds do not interact for the field range under consideration). Values of $D$ = 1258 and $E$ = -14 MHz, obtained from fitting $z \parallel B_0$ and $x \parallel B_0$, were used to calculate the transitions. The values of $D$ and $E$ differ only slightly from the ZFS parameters extracted from the monomer triplet spectrum.

As with the crystalline powder spectra, the features associated with $^5TT$ can be roughly distinguished from those of T+T, if they are detected at early times (Figure 4a) but not at late times (Figure 4b). Near $z \parallel B_0$ and $x \parallel B_0$, the two transitions from $^5TT_0$ are easily observable as pairs of absorptive/emissive inner peaks. Intermediate orientations (e.g., 50° and 120°) have a large degree of expected overlap between transitions, but it seems that the $^5TT$ peaks are generally less prominent. The spectra at later times (Figure 4b) are far simpler and feature no more than two peaks, which align well with the calculated transitions for T+T. These observations confirm the highly oriented nature of the single crystal.

The 25 – 75 ns spectra and the initial populations of the $^5$TT and T+T sublevels were calculated at all orientations with the parallel *JDE* model as described in Methods. This procedure simulates the spectra at $z \parallel B_0$ (ignoring broad inner peaks) and $x \parallel B_0$ with an exceptional degree of accuracy.(6) Simulations of the spectra at intermediate orientations were also successful (Figures S11-S13), especially regarding the predicted population of T+T. As with the crystalline powder spectra, many spectra exhibit a trend towards preferential population of the lower energy $M_S$ sublevels. Minor discrepancies between the simulation and data are evident both in peak position and amplitude and are likely related to the simplicity of the model—the spectra were simulated with only two adjustable parameters.

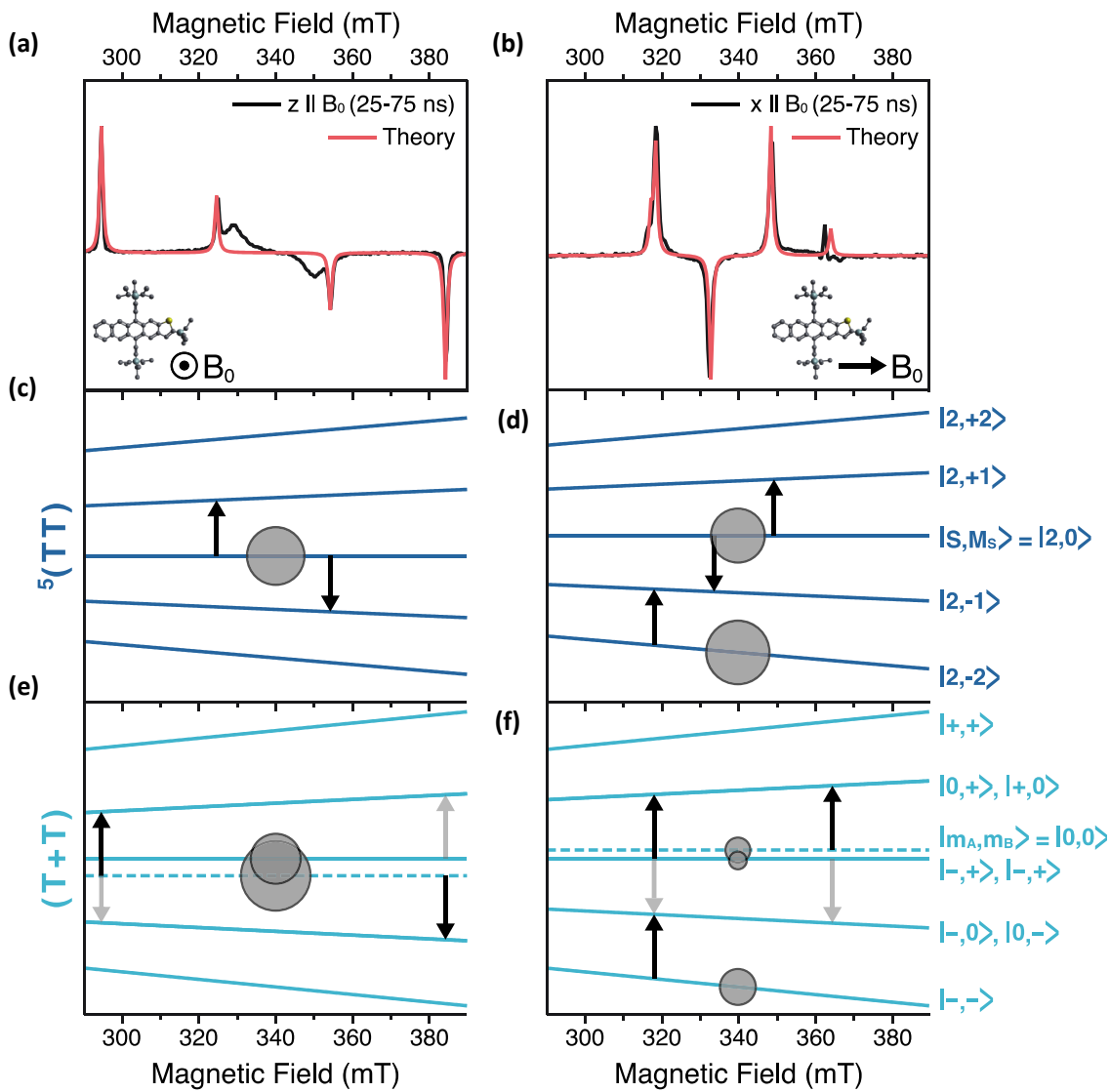

*Figure 5.* Early time (25 -75 ns) spectra and calculations of single crystal sample. Spectra are shown for (a) z' || $B_0$ and (b) x' || $B_0$ with associated energy level plots for (c,d) $^5$TT and (e,f) dissociated triplets T+T. Based on the theory, the arrows indicate field position of relevant transitions, with the associated circle areas indicating relative populations of the relevant sublevel.

The change in $^5$TT and T+T spectral amplitudes vs. delay time can be most clearly discerned at $z \parallel B_0$ (Figure 5a), where $^1$TT only relaxes into $^5$TT$_0$, and there is no overlap between the transitions from this species and T+T. Kinetic fitting at $\theta = 0°$ (Figure 6, Table 2) proceeded with a 20 ns resonator response function and an exponential rise and decay, with an extended (>10 µs) decay applied to account for a small (< 5%) population of long-lived T+T. The sharp peak associated with $^5$TT$_0$ → $^5$TT$_{+1}$ rises with $\tau$ = 28 ns, whereas the $|0\rangle \rightarrow |+\rangle$ peak associated with T+T appears with $\tau$ = 77 ns. However, close examination of the $|0\rangle \rightarrow |+\rangle$ kinetics reveals that the fit does not capture a portion of early signal amplitude. In line with this observation, no spectral traces were obtained (even within the resonator response time) in which $^5$TT was observed without T+T. The same fitting procedure was also applied to $\theta$ = 50 and 90° (Figure S14). TR-EPR spectra exhibit a significant degree of overlap at intermediate orientations (Figure 4), but field points could be found that lead to sliced kinetics with negligible population after 0.5 µs (considered predominantly $^5$TT). Comparison of the $^5$TT kinetics at different orientations (Table 2) reveals the trend that $^5$TT peaks persist longest at $x \parallel B_0$, second longest at $z \parallel B_0$, and decay noticeably faster at intermediate orientations.

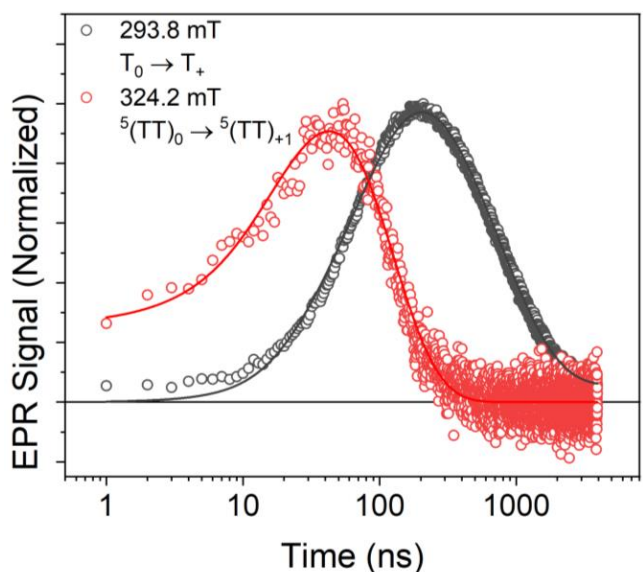

**Figure 6.** *Normalized kinetics at field positions where the $^5TT_0 \to {}^5TT_{+1}$ (red) and free triplet $T_0 \to T_+$ (dark gray) transitions are expected to occur at $z \parallel B_0$ ($\theta = 0°$).*

**Table 2.** Time constants of the rise and decay of representative peaks of $^5TT$ and T+T at select orientations using a single exponential rise and decay model.

| Orientation | $\tau_{rise}(^5TT)$ [a] | $\tau_{decay}(^5TT)$ [a] | $\tau_{decay}(T+T)$ [b] |
|---|---|---|---|
| 0° ($z \parallel B_0$) [a] | 28 ± 2 ns | 86 ± 3 ns | 735 ± 2 ns |
| 50° [b] | 14 ± 5 ns | 68 ± 6 ns | 1211 ± 5 ns |
| 90° ($x \parallel B_0$) [c] | 16 ± 1 ns | 100 ± 1 ns | 990 ± 3 ns |

[a]At $B_0$ = 324, 318, 340 mT for 0, 50, 90°, respectively. [b]At $B_0$ = 294, 348, 333 mT for 0, 50, 90°, respectively.

**Discussion**

**Magnetic field effects.** The observation of only two dips in emission intensity vs. magnetic field (Figure 3) in TES TIPS-TT provides further evidence of favorable molecular alignment, unlike lower symmetry TIPS tetracene samples (Figure S9).(17, 18) The presence of the third dip for TIPS tetracene at magnetic fields associated with $^1TT$ / $^3TT$ mixing could result from the lack of inversion center between molecular pairs and weak spin-orbit interactions. Antisymmetric spin-spin interactions,(27, 28) whose magnitude may be estimated by ($\Delta g/g$) $J$, where $g$ is the triplet $g$-factor and $\Delta g = g - g_e$, could drive such mixing. Several relevant TES TIPS-TT dimers possess inversion symmetry (Figure 2), negating this effect, while others (e.g., dimer Ia) possess a *J* much smaller than that of TIPS-tetracene that likely renders antisymmetric spin-spin interactions negligible.

The additional prominent feature that occurs at $B_0$ < 0.1 T in both TES TIPS-TT and TIPS tetracene is ubiquitous in early magnetic-field dependent experiments(29) on crystalline acenes. For triplet pairs with *J* = 0, states with "singlet character" mix at fields strengths similar to the zero-field splitting interaction. Its presence here alongside features associated with |*J*| = 15.4 GHz affirms that both paired (|*J* | >> |*D*|) and unpaired (*J* = 0) triplets exist in the crystal. The potential for fast TT dissociation, described below, rationalizes the detection of both species in time-integrated experiments.

We note that the multitude of possible molecular pairs will lead to other values of |*J*|; however, these may not be observable in the magneto-photoluminescence experiment for various reasons: triplet-triplet interactions on these pairs may not lead to detectable fluorescence that reflects the $^1$TT population, or they may be too weak or too strong to be detectable in the magnetic field range of the experiment. DFT calculations indicate *J* = -15 GHz for dimer Ia, making it the site most likely responsible for the observed resonant changes in PL. Based on the calculated value for dimer II (*J* = -315 GHz), dips would be expected at 16.9 and 33.8 T, well outside the range of the current experiment. However, signals for dimer Ib (DFT calc. *J* = 35 GHz) would be expected at 1.9 and 3.8 T but are not observed.

**Model rationale and limitations**. The highly successful fits of sharp TR-EPR features at all TES TIPS-TT crystal orientations underscore the successful convergence of parallel intermolecular orientation, macroscopic crystal alignment, and rigorous theory.(6) A critical result of the NTT presented in this work and in Ref. 6 is that the populations of the *M*-spin sublevels depend on the orientation of the chromophore pair relative to the magnetic field. The orientational dependence for sublevel populations is a prediction that is distinct from other treatments in the literature. Figure 5 shows that the populations in the $^5TT_M$ sublevels do indeed depend on orientation. But we also show that the populations of the T+T levels also depend on orientation—the quantum coherence imprinted on the singlet $^1TT$ state from singlet fission leads to distinct, and measurable, polarizations in both the TT and the unpaired T+T spectra. Because the unpaired singlet state $^1TT \rightarrow$ T+T does not have an EPR spectrum (coefficients of $|00\rangle$, $|+-\rangle$ and $|-+\rangle$ components of $^1TT$ are equal), the assignments in Figure 5 show that the T+T spectrum is from an unpaired quintet: $^1TT \rightarrow {^5TT} \rightarrow$ T+T.

As we are concerned with potential of specific spin sublevels of TT for QI purposes, it is imperative to distinguish pure from mixed states, and this distinction renders the JDE model more appropriate than Merrifield's theory of triplet (pair) populations,(29) based on triplet-triplet annihilation. First, the putative unpaired T+T "state" in the Merrifield theory is not a pure quantum state but rather a mixed state with a density matrix, but not a wavefunction. The literature concerning the formation of this state is somewhat murky, sometimes invoking states like "(T…T)" that may or may not be pure quantum states. For fitting TR-EPR data, the nature of the intermediate states—pure or mixed—may not be of much interest, but in quantum applications it is crucial. Secondly, the "singlet character" approximation in the original theory and oft-resurrected in recent literature(20, 30) resembles a Franck-Condon approximation, but it is unclear why such an approximation should be valid for the triplet-pair EPR spectra from singlet fission. Lastly, and perhaps most obviously, the Merrifield theory does not consider the potentially dominant *J* interaction between chromophores that eliminates essential curve-crossings in many systems.

The simple view of the dynamics we employ, which evolves the $^5$TT populations instantaneously, cannot be expected to be valid on all timescales. But the results we show here indicate that it is sufficient to capture many important features in the observed TR-EPR spectra, and it provides microscopic insight into the unpairing process with a minimal number of empirical parameters. Below we discuss aspects of the TES TIPS-TT TR-EPR spectra that fall outside the scope of the *JDE* model and speculate on their fundamental origins.

**Spin sublevel populations.** An explanation for the absence of $^5$TT$_{+M}$ population is provided by the value |*J*| = 15.4 GHz from Figure 3a, which leads to a crossing between the $^5$TT and $^3$TT manifolds, provided *J* < 0, within the magnetic field range probed by the X-band EPR experiment (300 – 400 mT, Figure 3b, Figure S15). Based on experimental results obtained by Chen et al.,(30) the $^5$TT$_{+2}$ / $^3$TT$_{-1}$ level crossing dictated by ferromagnetic *J* = -15.4 GHz may allow for transfer of $^5$TT$_{+2}$ population to $^3$TT$_{-1}$. We posit that the formally forbidden interaction of these states becomes allowed due to minor deviations (static or dynamic) from pure parallel symmetry for the TES TIPS-TT dimers in the crystal, which would enable $^5$TT / $^3$TT population transfer (but, we emphasize, not $^1$TT / $^3$TT mixing without additional perturbations).(31)   If TT hops between different sites during the early times of the TR-EPR experiment, a large portion of triplet pairs will at some point reside on the *J* = -15.4 GHz site within the instrument response time, allowing $^5$TT$_{+2}$ → $^3$TT$_{-1}$ transfer to occur specifically at this site. Chen *et. al.* proposed that the $^3$TT manifold enables fast, spin-allowed annihilation of one of the triplets: $^3$TT → $^3$(T + S$_0$).(30) Their mechanism would suggest that $^5$TT$_{+2}$ → $^3$TT$_{-1}$ transfer is the first step in an irreversible process that gradually filters out all exciton pairs from the $^5$TT$_{+2}$ state, which would

be reflected in the EPR spectra even if not all TT are occupying the $J$ = -15 GHz site at the time of spin-state measurement.

The precise field position of the crossing will depend on the orientation, but, as $J$ fluctuations at specific sites are believed to follow large normal distributions at room temperature,(32) a high likelihood of transient $^5TT_{+2}$ / $^3TT_{-1}$ level crossings can reasonably be assumed at all field points and orientations probed here. Larger $J$ fluctuations would be required to explain the depletion of $^5(TT)_{+1}$, and this brings forth the question about how exclusive the applied field-induced level crossing mechanism might be. We note the abundance of primarily absorptive TR-EPR spectra reported for different molecular systems(24-26) that may point towards the existence of a more general mechanism leading to inhibited population of $^5TT_{+1,+2}$, such as the one outlined by Nagashima et al. for TIPS-pentacene solids.(26) A theoretical approach to the problem of asymmetric population transfer for a TES TIPS-TT crystal requires further investigation, but it seems likely that the direct influence of the $|J|$ = 15.4 GHz site is a crucial piece of the puzzle.

**Shifting and line broadening of TR-EPR transitions.** The broad features in the $z \parallel B_0$ spectra are much better described by a Lorentzian than a Gaussian lineshape (Figure 7), indicating lifetime broadening vs. inhomogeneous broadening that might otherwise dominate a long-lived species in a molecular solid.(33, 34) The presence of two sets of $^5TT_0$ peaks is therefore attributed to a difference in average TT encounter times at distinct dimers, with TT at some sites being so short lived that their linewidths are dictated by lifetime broadening. The simplest analysis assuming a pure Lorentzian lineshape suggests a lifetime of roughly 0.9 ns for the broad inner peaks vs. ≥ 5 ns for the sharp outer peaks.

Although the proposed effect resembles the familiar exchange broadening in EPR,(35) wherein spins exchange quickly between two magnetically distinct sites, its relationship to triplet pairs is tenuous. Here, the interconversion of $^5$TT $\leftrightharpoons$ T+T reflects a fundamental change from $S = 2$ to $S = 1$ rather than the commonly analyzed case of a spin species (e.g., a hydrogen nuclei or radicals) merely undergoing a change in environment. However, our observations of broadening are consistent with TR-EPR measurements for which frequencies of interactions are comparable to the timescale of the experiment.(36) The details of this apparently unique form of broadening are worthy of further investigation.

The inter-chromophore anisotropic interaction $X$, which only affects TT and is dependent on intermolecular distance and orientation, was not included in the simulations presented thus far. In the absence of $X$, and in the Zeeman basis, transitions involving $^5$TT$_{\pm 2}$ overlap perfectly with those of T+T. Consequently, contributions from $X$ may explain the presence of small, relatively sharp side peaks neighboring T+T transitions at early times, as observed in the spectra at certain orientations (e.g., Figure 4a at 20°). The likelihood is small that $X$ may also be directly responsible for the broad but prominent peaks at $z \parallel B_0$ that overlap sharp transitions from $^5TT_0$. An experimental determination of the magnitude of $X$, currently unavailable for TES TIPS-TT, would help to uncover its role in the TR-EPR spectra.

**Exciton pair dynamics.** The higher $J$ calculated for dimer II can largely be attributed to the closer intermolecular distance (7), which places it in a favored position for fast triplet pair formation.(37) However, this initial population imbalance at the highest $J$ site must quickly evolve in order to explain the absorptive character of the TR-EPR spectra tentatively caused by $^5TT$ / $^3TT$ crossing at the $J$ = -15 GHz sites (dimer Ia), observable even in the first 20 ns. This necessary early evolution supports the notion that, while the $^5TT$ signal is observed for $\tau \approx 100$ ns, TT remains at a specific site for no more than several ns, and therefore the TR-EPR spectra reflect a shifting equilibrium of TT and dissociated T+T, residing on multiple dimer types. Although the inequality in calculated $J$ suggests that dimers Ia and Ib may behave differently, we do not distinguish between them for the duration of this discussion, as the geometries are nearly identical and the calculated $J$ = 35 GHz for dimer Ib suggests no impact on the TR-EPR spectra via additional level crossings (Figure S16).

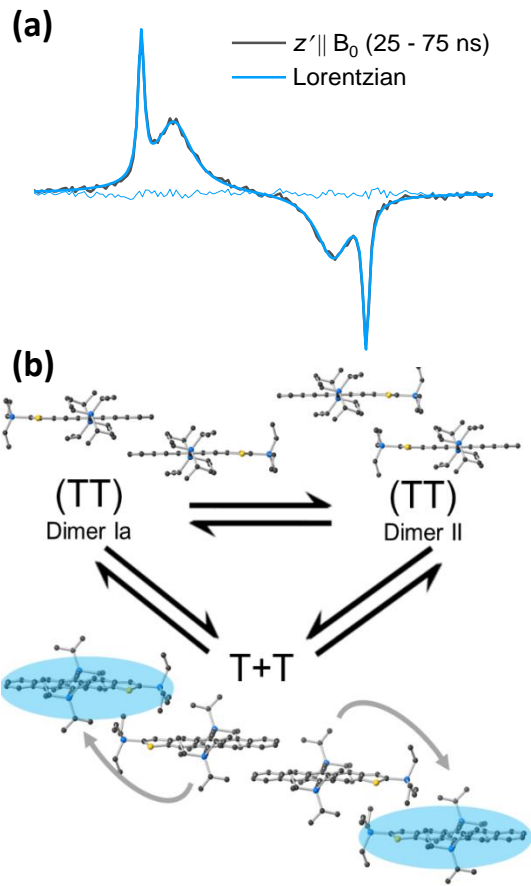

*Figure 7. Dynamic picture of triplet-pair populations. (a), Early time TR-EPR spectrum and fit to Lorentzian lineshape, characteristic of two types of lifetime-broadened quintets. (b), Depiction of equilibrium between distinct TT sites and dissociated T+T, also highlighted within the TES TIPS-TT crystal structure. Gray arrows and blue ovals indicate one possible route toward (T+T) formation from dimer II (TT).*

For a molecular crystal with extended order, there are multiple processes that would affect the lifetime of a particular TT species, e.g., triplet-triplet fusion or TT migration as observed in single crystals of tetracene.(38) However, among these processes, the dissociation into T+T involving proximal chromophores has been studied most extensively and likely provides justification for why different TT pair lifetimes, inferred from line broadening, are found within TES TIPS-TT. Triplet-pair dissociation has often been modelled using the so-called "transfer integral" approach,(39) which estimates coupling energies that facilitate the hopping of one member of TT to a neighboring chromophore. Single-exciton hopping rate constants for specific systems, such as TIPS pentacene, are derived from optical measurements involving bimolecular decay,(40-42) revealing that varying electronic coupling can lead to sub-ns hopping or make it prohibitively slow compared to unimolecular triplet decay.(24)

When considering dimer Ia, the relevant pair for the transfer integral is dimer II because the outlet for triplet pair dissociation involves the proximal molecules that form the phenyl-overlapped, vs. the thiophene-overlapped, geometry (Figure 7b). As has been shown theoretically,(43) slippage both along and perpendicular to the molecular long axis results in a modulation of both $J$ and the transfer integral that can lead to displaced maxima for these quantities. As such, the transfer integral for TT dissociation at dimer Ia could be relatively small, leading to a persistence of the exchange-coupled triplet pair at that site. Conversely, fast dissociation of the triplet pair at dimer II could be facilitated by a large transfer integral associated with thiophene-overlapped molecules (geometry of dimer Ia/b), and the broad inner quintet peaks are reflective of their fleeting population. A reliable transfer integral calculation is beyond the scope of this manuscript, but we note that with average triplet diffusion coefficients around $10^{-3}$ cm$^2$/s,(44) hopping along the c-axis (unit cell parameter ~ 21 Å), which is approximately parallel to the π-stacked staircase, would occur with a rate constant of roughly $2\times10^8$ s$^{-1}$. Further, the short-axis slippage of dimer II could lead to a slower rate constant of triplet motion that keeps the triplet pair at dimer Ia intact for longer than triplet pairs at dimer II.

Given that $^5$TT and T+T are present from the earliest TR-EPR observation times, we propose that the < 100 ns delay signals are representative of the equilibrium $^5$TT ⇋ T+T, with the narrow and broad $^5$TT peaks associated with the $J$ = -15 GHz and $J$ = -315 GHz sites, respectively. The disappearance of $^5$TT features is associated with a loss of spin coherence and decrease in exciton density due to population decay and eventual hopping orthogonal to the fastest transport direction, leading to fewer opportunities to re-fuse and generate $^5$TT. The different decay rates of the $^5$TT peaks (Table 2) suggests that the equilibrium decay may also be faster at intermediate orientations.(45, 46)  We also note that the degree of broadening exhibits some orientation dependence (Figure 5, compare the $^5$TT$_0$ transition peaks in $z \parallel B_0$ to $x \parallel B_0$). This may be caused by effects related to anisotropic hopping rates of triplets along different crystal directions or orientation-dependent spin-state interconversion.(42, 47)  Although we have not directly measured exciton transport for TES TIPS-TT, in rubrene a transition from one-dimensional to multi-dimensional triplet diffusion occurs on the 100s of ns timescale,(48) similar to $^5$TT decay times observed here and implying a transport-related mechanism of triplet-pair population loss.

**Conclusions.** We have presented a comprehensive analysis of the spin dynamics of triplet pairs in the novel organic compound TES TIPS-TT, which forms crystals that enforce molecular alignment that enables selective transfer of $^1$TT population to $^5$TT$_0$ when properly oriented in a magnetic field. Crystals of TES TIPS-TT also exhibit a strong modulation of fluorescence at high magnetic field, which allows for direct measurement of the exchange coupling between triplet pairs harbored on at least one nearest-neighbor dimer in the crystal lattice. By comparing TR-EPR spectra at various orientations with respect to the applied magnetic field with calculations using the *JDE* model, various aspects of the initial spin polarization have been elucidated. Control of spin sublevel population in this fashion advances toward the goal of harnessing triplet pairs as optically driven elements of quantum information processing.

**Materials and Methods**

**Experimental Details.**

**Structural characterization.**

X-ray diffraction data were collected at 90.0(2) K on a Bruker D8 Venture dual-source diffractometer with graded-multilayer focused MoK(alpha) X-rays. Raw data were integrated, scaled, merged, and corrected for Lorentz-polarization effects using the APEX3 package.(49) Corrections for absorption were applied using SADABS.(50) The structures were solved by dual-space methods (SHELXT)(51) and refined against $F^2$ by weighted full-matrix least-squares (SHELXL-2018).(52) Hydrogen atoms were found in difference maps but subsequently placed at calculated positions and refined using riding models. Non-hydrogen atoms were refined with anisotropic displacement parameters. The final structure model was checked using established methods.(53, 54) Atomic scattering factors were taken from the International Tables for Crystallography.(55) Additional crystal data and information on structure refinement can be found in the SI.

**Magnetophotoluminescence.** Column-shaped TES TIPS-TT crystals were attached to glass substrates using silver adhesive and then positioned at the face of a 50-micron diameter multi-mode optical fiber. The fiber was one port of a 50:50 coupler linking the sample with the excitation and detection arms. The excitation source was a filtered 519-nm diode laser operated below threshold giving an unpolarized power of 1 µW exiting the sample arm. Photoluminescence collected there reached the detection arm where it was coupled through a 539-nm edge filter and into a 0.27-m spectrometer with a cooled charged-coupled device array. The sample arm was held in He vapor within a 2 K, 0-14 T magnet (Quantum Designs, Dynacool) with the field oriented perpendicular to the long axis of the crystal.

**Electron paramagnetic resonance spectroscopy.** TR-EPR experiments at X-band (~9.5 GHz) were performed using a Bruker Elexsys E-580 spectrometer equipped with an ER 4118X-MS3 resonator. Spectra were collected after photoexcitation with 7 ns, ~2.5 mJ pulses from an Opotek Radiant 355 LD laser system under constant irradiation with microwave power of 2.4 mW. The quality factor of the resonator was measured to be between 500 – 700 for experiments on single crystal samples for which the kinetics were analyzed, indicating an expected resonator response function between 16 and 23 ns. Based on the average of these values, a resonator response of 20 ns was used for kinetic fitting.

The TIPS TES-TT monomer triplet spectrum was collected from a sample consisting of the material dissolved in a 4:1 mixture of iodobutane and toluene prepared in the glovebox, temporarily sealed with a septum, frozen using liquid $N_2$, and then rapidly transferred to the EPR spectrometer held at 100 K to prevent oxygen from dissolving into the solvent matrix of the sample. The crystalline powder sample was prepared by placing glass capillaries coated with small amounts of crystalline powder into clear fused quartz (CFQ) EPR tubes, which were then flame-sealed under vacuum. Four single crystal samples were mounted using the (0 1 -1) or (0 -1 1) faces to the end of CFQ rods cut at 38° in a manner shown in photos provided in the SI. Using a goniometer manufactured by Bruker, multiple spectra were taken for each sample to determine the orientation (within 2°) at which the splitting between the two peaks assigned to diffuse triplets was the largest. As the largest splitting between triplet transitions is expected to occur when the primary molecular z-axis is parallel to the applied magnetic field, the sample exhibiting the largest splitting was chosen as the most well-aligned and was used for further analysis.

**Theoretical Development and Calculations**

***Theoretical Development.*** To model the exciton unbinding process, we let the *J*-coupling be binary—it is *J* when two molecules in the crystal are nearest neighbors and zero otherwise. The Hamiltonian takes the form,

$$\mathcal{H} = H_{Zeeman,AB} + H_{ZFS,A} + H_{ZFS,B} + f(J\vec{S}_A \cdot \vec{S}_B), \qquad (1)$$

for chromophores *A* and *B* where *f* is a binary switching function that is either zero or one. While the $^1$TT state forms on adjacent chromophores,(8) the excitons in a crystal are mobile. Once one of the excitons hops to another chromophore, the $^1$TT state can decohere and evolve to separated triplet pairs, T+T.

Part of the spectrum comes from the $^5$TT$_M$ sublevels, whose calculations appear in the analysis of the parallel *JDE* model in an earlier publication.(6) But in a crystal, some of the triplets in the ensemble will have hopped and unpaired, even at early times. We assume that the jump dynamics are slow relative to the timescale for quintet formation on neighboring molecules but fast enough to completely dephase the triplet pair states. We model this by allowing the quintet state to form for all molecules in the *JDE* Hamiltonian, $f = 1$. Some fraction of those molecules in the ensemble will experience a jump between time zero and time *t*. That sub-ensemble will quench into the unpaired triplet Hamiltonian, $f = 0$. We quench those molecules, using the density matrix from the initial quintet states of the *JDE* model, into the states of the unpaired triplet by applying the projection operator $P = \sum_{M_A,M_B} |M_A, M_B\rangle \langle M_A, M_B|$, where the sum over $M_A$ and $M_B$ goes over the sublevels of unpaired states—the eigenstates of the Hamiltonian with $f = 0$. The projection operator separates the diagonal elements of the density matrix in the unpaired basis. The density matrix of the entire system is $\rho = w\rho_0 + (1-w)P\rho_0 P$, where $w$ is an empirical parameter equivalent to the fraction of exciton pairs that have not undergone a jump before the time of measurement, and $\rho_0$ is the density matrix of the *JDE* model at early times.(6) Here, we are assuming that the populations decouple from the coherences and that the time evolution of the coherences is fast compared to the populations. Both approximations appear in the Redfield theory of quantum relaxation.(56) The calculation of the TR-EPR spectrum follows from the density matrix.(6)

*TES TIPS-TT TR-EPR calculations*. The single crystal spectra of TES TIPS-TT at select orientations were calculated with the parallel *JDE* model, where both transitions and intensities are calculated from the spin Hamiltonian (Eq. 1). The computed TR-EPR spectra at each orientation are sums of two components: one from the $^5TT_M$ sublevels and another from the spin-polarized, unpaired triplets T+T. First, the calculated spectrum was compared to the data for *z* ∥ $B_0$ to optimize a value of *D* in the least-squares sense with the simulated annealing technique. At this orientation, *E* has little to no effect on the spectrum. Fixing *D* to the resulting best-fit value (1260 MHz), the *x* ∥ $B_0$ spectrum was optimized for *E* (-16 MHz). When optimizing parameters, the Hamiltonian is evaluated in the eigenbasis of the quintet subspace, which we call the adiabatic basis, $|S=2, \alpha\rangle = \sum_M \alpha_M |S=2, M\rangle$. These states are very close to the Zeeman $|S=2, M\rangle$ states *away from crossings*.(31) Although the energies of the Zeeman states change upon sample rotation, the states remain well defined. The $^5$TT and T+T populations were calculated using the parallel *JDE* model (see above), but to replicate the data, populations of the two high energy quintet states ($M = +1, +2$) were set to zero. The four $^5$TT and two T+T lines ($\Delta M = \pm 1$) at each orientation are broadened by Lorentzian lineshapes. Line intensities are proportional to the difference in population between the $\Delta M = \pm 1$ sublevels and the corresponding dipole matrix element squared. The relative amplitudes of the $^5$TT and T+T spectra are orientation-dependent and are estimated from the data. Diagrams of the spin sublevel energies and populations (Figures 5c-f, S11-S13) were likewise calculated, but in the diabatic Zeeman basis where states of *S* and *M* are long-lived.

***Monomer $^{3*}$(TES TIPS-TT) TR-EPR calculations.*** The monomer $^{3*}$(TES TIPS-TT) powder EPR spectrum was calculated by evaluating the spin Hamiltonian $\mathcal{H} = H_{Zeeman} + H_{ZFS}$ in the single triplet exciton (T) eigenbasis. The simulated annealing optimization method found the best-fit parameters D = 1273 MHz and E = -40 MHz. Intersystem crossing populates the zero-field triplet states: $|x\rangle$, $|y\rangle$, and $|z\rangle$. Because the nonadiabatic transition theory in the *JDE* model is for triplet *pairs*, the zero-field populations are fit parameters ($p_x = 0.11$, $p_y = 0.11$, and $p_z = 0.78$). To simulate the powder EPR spectrum, Hamiltonians were calculated for a spherical distribution of orientations describing the relative orientation of the chromophore with respect to the field. An additional geometrical factor was applied to the amplitude of the spectrum from each orientation before summing.

**Acknowledgments**

The authors thank Obadiah Reid and Ryan Dill for useful discussions and Petri Alahuhta for assistance with sample preparation. This work was authored by Alliance for Sustainable Energy,

LLC, the manager and operator of the National Renewable Energy Laboratory for the U.S. Department of Energy (DOE) under Contract No. DE-AC36-08GO28308. Funding provided by U.S. Department of Energy, Office of Basic Energy Sciences, Division of Chemical Sciences, Biosciences, and Geosciences.  A portion of the research was performed using computational resources sponsored by the Department of Energy's Office of Energy Efficiency and Renewable Energy and located at the National Renewable Energy Laboratory. This research also used resources of the National Energy Research Scientific Computing Center, a DOE Office of Science User Facility supported by the Office of Science of the U.S. Department of Energy under Contract No. DE-AC02-05CH11231. The views expressed in the article do not necessarily represent the views of the DOE or the U.S. Government. The U.S. Government retains and the publisher, by accepting the article for publication, acknowledges that the U.S. Government retains a nonexclusive, paid-up, irrevocable, worldwide license to publish or reproduce the published form of this work, or allow others to do so, for U.S. Government purposes.

38. Y. Wan, G.P. Wiederrecht, R.D. Schaller, J.C. Johnson, & L. Huang Transport of spin-entangled triplet excitons generated by singlet fission. *The Journal of Physical Chemistry Letters* 9, 6731-6738 (2018).
39. J. Wehner & B. Baumeier Intermolecular singlet and triplet exciton transfer integrals from many-body green's functions theory. *Journal of Chemical Theory and Computation* 13, 1584-1594 (2017).
40. C. Grieco*, et al.* Triplet transfer mediates triplet pair separation during singlet fission in 6,13-bis(triisopropylsilylethynyl)-pentacene. *Advanced Functional Materials* 27, 1703929 (2017).
41. G.S. Doucette*, et al.* Tuning triplet-pair separation versus relaxation using a diamond anvil cell. *Cell Reports Physical Science* 1, 100005 (2020).
42. R.J. Hudson, D.M. Huang, & T.W. Kee Anisotropic triplet exciton diffusion in crystalline functionalized pentacene. *The Journal of Physical Chemistry C* 124, 23541-23550 (2020).
43. V. Abraham & N.J. Mayhall Revealing the contest between triplet–triplet exchange and triplet–triplet energy transfer coupling in correlated triplet pair states in singlet fission. *The Journal of Physical Chemistry Letters* 12, 10505-10514 (2021).
44. T. Zhu, Y. Wan, Z. Guo, J. Johnson, & L. Huang Two birds with one stone: Tailoring singlet fission for both triplet yield and exciton diffusion length. *Advanced Materials* 28, 7539-7547 (2016).
45. A. Schweiger & G. Jeschke (2001) *Principles of pulse electron paramagnetic resonance* (Oxford University Press) 1 Ed.
46. N.F. Berk, J. Rosenthal, & L. Yarmus Spin relaxation of triplet excitons in molecular crystals. *Physical Review B* 28, 4963-4969 (1983).
47. W. Bizzaro, J. Rosenthal, N.F. Berk, & L. Yarmus The epr linewidth of triplet excitons in single-crystal anthracene: A re-examination. 84, 27-32 (1977).
48. E.A. Wolf & I. Biaggio Geminate exciton fusion fluorescence as a probe of triplet exciton transport after singlet fission. *Physical Review B* 103, L201201 (2021).
49. B.-A. APEX3 (2018).
50. L. Krause, R. Herbst-Irmer, G.M. Sheldrick, & D. Stalke Comparison of silver and molybdenum microfocus x-ray sources for single-crystal structure determination. *Journal of Applied Crystallography* 48, 3-10 (2015).
51. G.M. Sheldrick Shelxt - integrated space-group and crystal-structure determination. *Acta Crystallographica Section A* 71, 3-8 (2015).
52. G.M. Sheldrick Crystal structure refinement with shelxl. *Acta Crystallographica Section C* 71, 3-8 (2015).
53. S. Parkin Expansion of scalar validation criteria to three dimensions: The r tensor. Erratum. *Acta Crystallographica Section A* 56, 317 (2000).
54. A.L. Spek Structure validation in chemical crystallography. *Acta Crystallographica Section D* 65, 148-155 (2009).
55. A.J.C. Wilson (1992) *International tables for crystallography. Volume c: Mathematical, physical and chemical tables* (Kluwer Academic Publishers, Holland).
56. K. Blum (2012) *Density matrix theory and applications* (Springer Science & Business Media).
32